\documentclass[10pt,conference]{IEEEtran} \IEEEoverridecommandlockouts
\usepackage{cite}
\usepackage{amsmath,amssymb,amsfonts}
\usepackage{algorithmic}
\usepackage{graphicx}
\usepackage{textcomp}
\usepackage{xcolor}
\usepackage{verbatim}
\usepackage{graphicx}
\usepackage{multirow}
\usepackage{tabularx}
\usepackage{booktabs} %
\usepackage{array}    %
\usepackage[margin=1in]{geometry} %
\usepackage{url}

\usepackage{authblk}

\def\BibTeX{{\rm B\kern-.05em{\sc i\kern-.025em b}\kern-.08em
    T\kern-.1667em\lower.7ex\hbox{E}\kern-.125emX}}
\begin{document}

\title{Benchmarking Large Language Models on Repairing Qiskit Programs using Bugs4Q}

\author[1,$\dagger$]{Saumya Brahmbhatt}
\author[1,$\dagger$]{Mitali Hukkeri}
\author[1]{Dongchan Kim}
\author[2]{M.V. Panduranga Rao}
\author[1]{Lei Zhang}

\affil[1]{Department of Information Systems, University of Maryland, Baltimore County, USA}
\affil[2]{Department of Computer Science and Engineering, Indian Institute of Technology Hyderabad, India}

\affil[1]{{\{saumyab1,mhukker1,dkim26,leizhang\}@umbc.edu}} 
\affil[2]{{\{mvp\}@cse.iith.ac.in}} 

\affil[$\dagger$]{These authors contributed equally to this work.}

\maketitle

\begin{abstract}

In quantum programs, Bugs4Q is a widely used benchmark containing real quantum defects. However, its evaluation assumes that benchmark labels remain valid and that generated fixes execute in the target environment. We evaluate two Bugs4Q versions containing 67 unique real Qiskit defects, adding executable tests where missing, and re-run all entries across six pinned Qiskit releases (0.25.0, 0.45.0, 1.0.0, 1.1.1, 2.0.0, and 2.3.1). We find that quantum benchmarks can suffer from silent label inversion: entries become invalid without errors when reference fixes stop executing or buggy programs no longer reproduce failures. Thus, correctness depends on the (benchmark, version) pair rather than the benchmark alone. We evaluate four LLMs (GPT-4o-mini, GPT-5o-mini, GPT-5.4, and GPT-5.4-mini), generating up to 10 repair candidates per defect and testing them across all versions. GPT-5.4 achieves the highest pass@10 (48.8\%), followed by GPT-5.4-mini (47.3\%), GPT-5o-mini (30.3\%), and GPT-4o-mini (22.6\%). All models perform best on Qiskit 0.45.0 and decline after the Qiskit 1.0 transition. Many failures arise from deprecated or incompatible APIs rather than incorrect repairs, and 64\% of successful repairs occur on entries invalid under the target version. We release a re-validated, version-pinned Bugs4Q benchmark and show that benchmark validation must precede repair evaluation.

\end{abstract}


\section{Introduction} 
While testing and debugging quantum software remain challenging due to quantum mechanics, such as superposition and entanglement\cite{preskill2018quantum,murillo2025quantum,miranskyy2021testing}, 
automated program repair (APR) utilizing large language models (LLMs) has significantly advanced this agenda in classical software. Prior work has shown that large pre-trained models can outperform earlier APR techniques under several settings~\cite{monperrus2018automatic,xia2023automated,jiang2023impact,jin2023inferfix}.

In contrast, LLM-based repair of quantum software is still in its early stages. Existing studies provide promising but fragmented evidence. Guo \emph{et al.} evaluated ChatGPT on Bugs4Q and showed that ChatGPT-based repair is feasible on real defects~\cite{guo2024repairing,zhao2021bugs4q}. QBugLM~\cite{pham2026qbuglm} studies quantum debugging with taxonomy-driven bug injection in OpenQASM 3.0. QuanBench+ evaluates multi-framework quantum code generation with executable tests and a feedback-based repair setting~\cite{slim2026quanbenchplus}. 
However, these APR studies rely on benchmarks whose labels and executability are stable. A recent reproducibility study showed that Bugs4Q degrades substantially across Qiskit versions, with reproducibility dropping sharply on modern releases~\cite{ohto2026reproducibility}. %

Motivated by this gap, we benchmark four contemporary LLMs (GPT-4o-mini, GPT-5o-mini, GPT-5.4, and GPT-5.4-mini) on repairing Qiskit programs using the two Bugs4Q releases with a version-pinned evaluation protocol. We further author executable tests for entries that originally lacked them, and we execute both benchmark references and model-generated fixes under six controlled Qiskit environments (0.25.0, 0.45.0, 1.0.0, 1.1.1, 2.0.0, and 2.3.1) spanning the major-version transition around Qiskit~1.0. Our goal is not only to measure how often LLMs repair real quantum defects, but also to determine how much of the reported success or failure is attributable to repair logic versus environmental compatibility.

This paper makes four major contributions. First, we construct a deduplicated benchmark of 67 real Qiskit defects from Bugs4Q-Framework~\cite{bugs4q-framework} and Bugs4Q-NA~\cite{bugs4q-na}, and we author executable test oracles for cases that originally lacked them. Second, we introduce a version-pinned validation protocol across six Qiskit releases and show that benchmark validity is a property of the (case, version) pair rather than the benchmark alone. Third, we evaluate four LLMs under this protocol and find that pass rates vary by more than a factor of two across versions for the same model, peaking at 0.45.0 and collapsing across the Qiskit~1.0 boundary. Finally, we demonstrate that environment incompatibility accounts for 13--56\% of repair failures depending on the model, and that 64\% of observed passes occur on invalid benchmark entries. %
We release the artifacts at \url{https://github.com/slrrla/APR-new-experiment/tree/APRFinal} for reproducibility.

\section{Our Method}\label{sec:methodology}

Our evaluation runs in five stages as detailed in the next subsections.

\subsection{Benchmark Construction}

We draw our benchmark from two sources: the original Bugs4Q framework~\cite{bugs4q-framework} and the more recent Bugs4Q-NA repository~\cite{bugs4q-na}, both of which collect real Qiskit bugs reported through GitHub, Stack Overflow, and Stack Exchange~\cite{zhao2021bugs4q}. After merging the two and discarding entries that pointed to the same underlying defect, we obtain 67 unique bug cases.

Every case comes with a buggy program, a reference fix, and a natural-language issue description. We keep the issue description because it is the only problem context a developer would realistically have when attempting the repair, and we hold the model to that same information: LLMs see the buggy program and the issue description, but never the reference fix or the test oracle.

\subsection{Test Construction}

For Bugs4Q entries, without executable tests, we created MUT-based oracles that accept a candidate repair through an environment variable and executable it with \texttt{runpy.run\_path}. Draft skeletons were generated with Claude Opus---a model family distinct from the GPT models under evaluation, which reduces oracle bias toward the repair outputs—and then revised by the authors. We accepted an oracle only if it failed on the buggy program, passed on the reference fix under the target version, and asserted on the behavior described in the issue rather than on incidental output; for probabilistic cases we pinned simulator seeds and shot counts where applicable. We also rewrote the 42 Bugs4Q-Framework tests to the same MUT interface as the 25 Bugs4Q-NA cases and, during validation, corrected 15 that used inverted assertions passing on the buggy behavior. Two authors independently reviewed all resulting tests against the buggy-fail/fixed-pass criterion and the reported issue intent.

\subsection{Version-Pinned Validation}

Qiskit's API has changed enough between releases that a case which behaves correctly under one version can stop reproducing the bug---or stop running at all---under another. We therefore validate every case under six pinned versions: 0.25.0, 0.45.0, 1.0.0, 1.1.1, 2.0.0, and 2.3.1. These span the pre- and post-1.0 environments and bracket the Qiskit~1.0 compatibility boundary.

For each case-version pair, we run both the buggy and the reference-fixed program against the bundled or authored test, and we mark the pair valid only when the buggy program fails and the fix passes under that same test. A buggy program that fails to execute under a given version counts as failing the test. When the buggy program no longer reproduces the reported fault, or the fix no longer runs, we drop the case from APR evaluation for that version.

\subsection{LLM-Based Repair Generation}

Repair generation and execution run on all case--version pairs; the validity matrix determines the denominators of the repair metrics we report. Each prompt gives the model the buggy source and the issue description, and nothing else: no reference fix, no test code, no execution logs, and no excerpts from Qiskit migration guides. The intent is to mimic a realistic repair setting in which the model has only the broken implementation and the bug report to work from.

We sample $n=10$ candidate repairs per (model, case, version). Each candidate replaces the buggy program and runs under the same pinned version used in validation; we count it as a success if it passes that case's test. 
\subsection{Evaluation Metrics and Failure Analysis}

Our primary metric is pass@$k$: the fraction of cases for which at least one of the first $k$ candidates (in generation order) passes the test under a given version, for $k \in \{1, 5, 10\}$. This empirical definition matches our fixed-order protocol; we do not use the combinatorial unbiased estimator, which assumes exchangeable samples.

Alongside pass@$k$ we track executability: whether a candidate parses, imports, and runs at all under the target version. When a candidate fails, the execution log distinguishes two outcomes---it does not execute (syntax errors, missing imports, deprecated or removed APIs), or it executes but fails the test. We keep these separate because they reflect different constructs: environment compatibility versus repair correctness.

Finally, we compare repair outcomes across versions. To keep the comparison fair, we restrict each one to cases that are valid in both versions being compared, so that a swing in pass@$k$ reflects the model rather than a case quietly dropping out of the benchmark between releases.

\section{Results}\label{sec:results}

\subsection{Benchmark validity depends on the Qiskit
version}\label{sec:res-validity}

\begin{table}[t]
\centering
\caption{Benchmark validation across pinned Qiskit versions.}
\label{tab:validation}
\small
\resizebox{\columnwidth}{!}{%
\begin{tabular}{lccccccc}
\toprule
Dataset & Cases & 0.25.0 & 0.45.0 & 1.0.0 & 1.1.1 & 2.0.0 & 2.3.1\\
\midrule
Bugs4Q-Framework & 42 & 30 & 23 & 9 & 9 & 5 & 5 \\
Bugs4Q-NA        & 25 & 6  & 6  & 3 & 3 & 5 & 2 \\
Deduplicated     & 67 & 36 & 29 & 12& 12& 10& 7 \\
\bottomrule
\end{tabular}
}
\end{table}

Table~\ref{tab:validation} reports, for each pinned release, how many of the 67 deduplicated cases remain valid: the buggy program still reproduces its fault, and the reference fix still executes and passes the associated test. Validity is not stable across releases: 63/67 entries fail to validate under at least one version, and only 4/67 are valid under all six pinned versions. Crucially, much of this invalidity is \emph{silent}. In 57 of the 402 (case, version) pairs, the buggy program passes the oracle outright---the reported fault no longer reproduces---including 54 pairs in which both the buggy and the fixed program pass, so the oracle cannot discriminate at all. In a further 105 pairs the reference fix executes cleanly yet fails its own test. Neither situation raises any error.

\subsection{Repair rates under version-pinned
evaluation}\label{sec:res-passk}

\begin{table}[t]
\centering
\caption{Repair pass@$k$ by Qiskit version (raw count out of 
67 cases per version, before validity filtering). The pre-1.0 
peak at 0.45.0 and the collapse across the Qiskit~1.0 boundary 
are consistent across every model.}
\label{tab:passk-version}
\small
\resizebox{\columnwidth}{!}{
\begin{tabular}{l ccc ccc ccc ccc}
\toprule
\multirow{2}{*}{Version} &
  \multicolumn{3}{c}{GPT-4o-mini} &
  \multicolumn{3}{c}{GPT-5o-mini} &
  \multicolumn{3}{c}{GPT-5.4} &
  \multicolumn{3}{c}{GPT-5.4-mini} \\
\cmidrule(lr){2-4}\cmidrule(lr){5-7}\cmidrule(lr){8-10}\cmidrule(lr){11-13}
 & @1 & @5 & @10 & @1 & @5 & @10 & @1 & @5 & @10 & @1 & @5 & @10 \\
\midrule
0.25.0 &  9 & 15 & 18 & 15 & 27 & 29 & 21 & 23 & 25 & 19 & 25 & 26 \\
0.45.0 & 15 & 24 & 29 & 22 & 35 & 39 & 36 & 44 & 46 & 34 & 39 & 43 \\
1.0.0  &  8 &  9 & 12 &  8 & 14 & 16 & 29 & 34 & 34 & 23 & 30 & 33 \\
1.1.1  &  8 &  9 & 12 &  9 & 15 & 17 & 29 & 34 & 34 & 23 & 30 & 33 \\
2.0.0  &  7 &  9 & 10 &  6 & 10 & 11 & 24 & 28 & 29 & 22 & 26 & 28 \\
2.3.1  &  7 &  9 & 10 &  4 &  9 & 10 & 17 & 18 & 28 & 16 & 19 & 27 \\
\bottomrule
\end{tabular}
}
\end{table}

\begin{figure}[t]
\centering
\includegraphics[width=0.8\columnwidth]{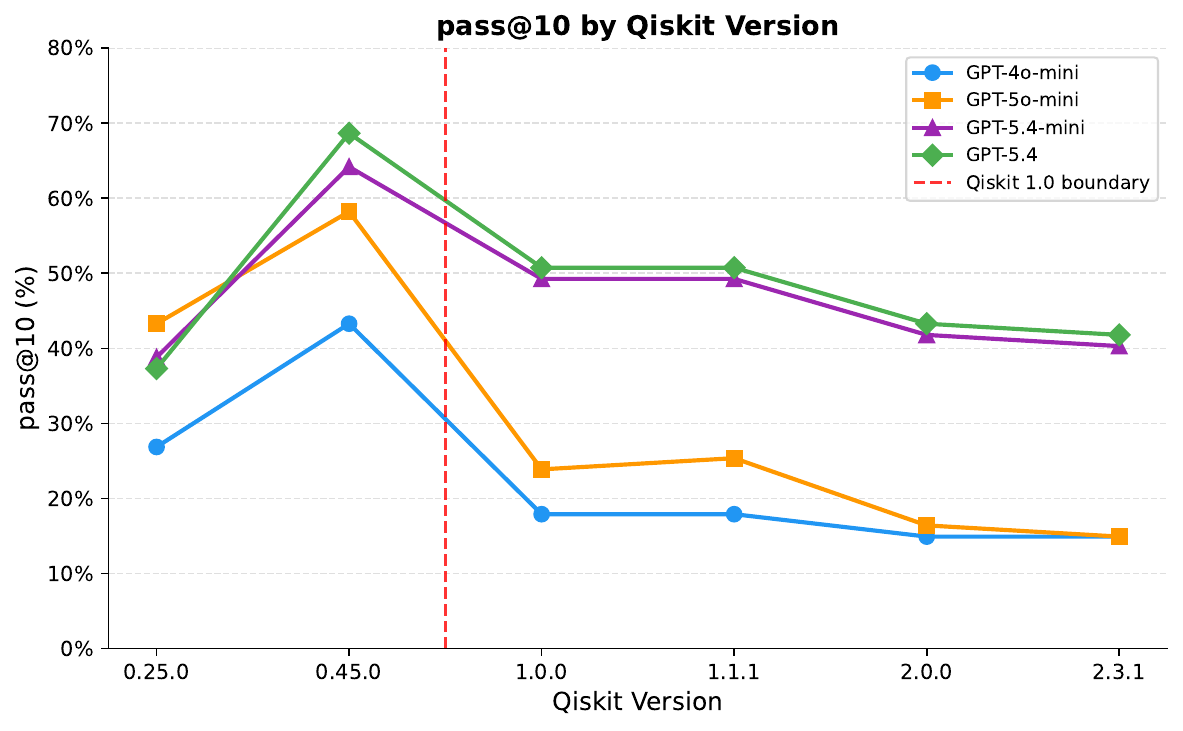}
\caption{pass@10 by Qiskit version for all four models. Every model peaks at
the pre-1.0 release 0.45.0 and drops sharply across the Qiskit~1.0 boundary
(dashed line), then continues to erode through the 2.x releases.}
\label{fig:passk-version}
\end{figure}

Table~\ref{tab:passk-version} and Fig.~\ref{fig:passk-version} report
unvalidated pass@$k$ for each model, computed over all 67 cases per version as a conventional evaluation would. Three patterns hold across all four models. First, repair rates are far from saturated even at pass@10: pooled across models, pass@10 never exceeds 58.6\% at any single version, and pass@1 never exceeds 39.9\%. Second, performance is strongly non-monotonic along the version axis, peaking at the pre-1.0 release 0.45.0 and declining on either side. Pooled pass@10 rises from 36.6\% at 0.25.0 to a maximum of 58.6\% at 0.45.0, then drops to 35.4\% at 1.0.0 and continues down to 28.0\% at 2.3.1; pass@1 follows the same shape (39.9\% at 0.45.0 versus 16.4\% at 2.3.1). Third, the single largest change occurs across the Qiskit~1.0 boundary: every model loses a substantial fraction of its passing cases between 0.45.0 and 1.0.0. GPT-5o-mini falls from 58.2\% to 23.9\% pass@10 and GPT-4o-mini from 43.3\% to 17.9\%, consistent with the removal of pre-1.0 interfaces at the major-version transition.

Model capability shifts the level of these curves but not their shape. The stronger GPT-5.4 and GPT-5.4-mini remain well above the smaller GPT-4o-mini and GPT-5o-mini at every version. GPT-5.4 attains the highest single-version result, 68.7\% pass@10 at 0.45.0, yet exhibits the same 0.45.0 peak and the same post-1.0 decline. The gap widens after the boundary: the smaller models retain only 17.9--23.9\% pass@10 at 1.0.0, whereas the GPT-5.4 pair retains roughly 49--51\%. A single version-agnostic number would obscure this entirely; for a fixed model, the same benchmark yields pass@10 values that differ by more than a factor of three across releases.

\begin{table}[t]
\centering
\caption{Validated pass@10: numerator and denominator restricted to
case--version pairs that survive validation (Table~\ref{tab:validation}).}
\label{tab:passk-validated}
\small
\resizebox{\columnwidth}{!}{
\begin{tabular}{lcccc}
\toprule
Version ($n$ valid) & 4o-mini & 5o-mini & 5.4 & 5.4-mini \\
\midrule
0.25.0 (36) & 11 & 19 & 14 & 13 \\
0.45.0 (29) &  7 & 17 & 19 & 16 \\
1.0.0 (12)  &  3 &  8 &  9 &  7 \\
1.1.1 (12)  &  3 &  8 &  9 &  7 \\
2.0.0 (10)  &  5 &  7 &  8 &  8 \\
2.3.1 (7)   &  2 &  5 &  5 &  5 \\
\bottomrule
\end{tabular}
}
\end{table}

Restricting the denominator to validated pairs changes the picture. Table~\ref{tab:passk-validated} reports pass@10 over the valid cases only. Validated repair rates do not collapse after Qiskit~1.0---GPT-5.4 passes 9/12 at 1.0.0 and 8/10 at 2.0.0---indicating that much of the post-1.0 collapse visible in Table~\ref{tab:passk-version} reflects the shrinking validity of the benchmark itself (36 valid cases at 0.25.0 versus 7 at 2.3.1) rather than declining model ability. The small validated denominators warrant caution---the Wilson 95\% intervals for the two rates above are 47--91\% and 49--94\% respectively, and the cases that remain valid on recent releases are plausibly the more robust ones---but the direction is consistent: version effects live primarily in the benchmark, not the model. For this reason we report validated results as raw counts rather than percentages. Most importantly, of the 599 passes observed across all models and versions, 384 (64.1\%) occur on case--version pairs that fail validation---passes that a conventional evaluation would report as repair successes even though the underlying benchmark entry is invalid.

\subsection{Failures reflect executability as much as repair
logic}\label{sec:res-exec}

The version structure of these results is itself evidence that a large share of apparent repair failures are environmental rather than logical. A candidate that passes under 0.45.0 but fails under 1.0.0 is not a less correct repair; it becomes non-executable under a stricter API. We quantify this fragility directly: of the 165 (model, case) pairs that pass at pass@10 under at least one version, only 40 (24.2\%) pass under all 6 versions, while 26 (15.8\%) pass under exactly one version. A majority of repair successes are therefore version-contingent, and the version on which a repair happens to be observed determines whether it is counted at all.

\begin{table}[t]
\centering
\caption{Best-of-10 outcome per case--version pair (402 per model). \emph{Fail} = executes but fails the test; \emph{Incompatible} = does not execute. Parentheses: incompatible share of non-passing candidates.}
\label{tab:outcomes}
\small
\resizebox{\columnwidth}{!}{
\begin{tabular}{lccc}
\toprule
Model & Pass & Fail (executes) & Incompatible \\
\midrule
GPT-4o-mini  &  91 & 178 & 133 (42.8\%) \\
GPT-5o-mini  & 122 & 124 & 156 (55.7\%) \\
GPT-5.4      & 196 & 180 &  26 (12.6\%) \\
GPT-5.4-mini & 190 & 175 &  37 (17.5\%) \\
\midrule
Pooled       & 599 & 657 & 352 (34.9\%) \\
\bottomrule
\end{tabular}
}
\end{table}

To attribute failures explicitly, we classify every non-passing candidate as either \emph{incompatible} (it does not execute---syntax or import errors, deprecated or removed APIs) or \emph{failing} (it executes but fails the test). Table~\ref{tab:outcomes} reports the distribution. Environment-related failures (non-executability and API incompatibility) dominate the smaller models: GPT-4o-mini has 42.8\% of its non-passing candidates classified as incompatible, and GPT-5o-mini reaches 86.1\%, whereas the stronger GPT-5.4 and GPT-5.4-mini show only 12.6\% and 17.5\% respectively, indicating their fixes are more version-robust but still frequently fail on repair logic. 

Finally, threats to validity are summarized in Table~\ref{tab:threats}.

\begin{table}[t]
\centering
\caption{Summary of threats to validity.}
\label{tab:threats}
\resizebox{\columnwidth}{!}{
\begin{tabular}{p{0.18\linewidth} p{0.76\linewidth}}
\toprule
\textbf{Threat} & \textbf{Summary} \\
\midrule
\textbf{Internal} &
Oracles were drafted with Claude Opus and reviewed by two authors, introducing reviewer dependence; using a model family disjoint from the evaluated GPT models mitigates but does not eliminate oracle--repair-model circularity. Weak or overly strict oracles may respectively inflate or deflate repair success.\\
\addlinespace
\textbf{Construct} &
Execution logs distinguish environment incompatibility from repair failures but cannot separate semantic correctness from test overfitting, so some passing repairs may reflect weak oracles. \\
\addlinespace
\textbf{External} &
Results are limited to Qiskit defects, four GPT-family models, and six pinned Qiskit versions; findings may not generalize to other frameworks, model families, or intermediate software versions. \\
\bottomrule
\end{tabular}
}
\end{table}

\section{Conclusions and Future Work}\label{sec:conclusion}

We benchmark 4 LLMs on repairing 67 real Qiskit defects from Bugs4Q across 6 pinned Qiskit versions. Our results show that benchmark validity is version-dependent, and that environment incompatibility accounts for a large share of apparent repair failuresWhile 64\% of observed passes occur on invalid benchmark entries, most repair successes are version-contingent, passing under only a subset of the six versions. %

These findings have an implication for how quantum APR results should be reported: a single version-agnostic pass rate conflates repair correctness with environment compatibility and is therefore not a reliable measure of model capability. We argue that version-pinned evaluation and explicit executability tracking should become standard practice in quantum APR benchmarking.

Future work should investigate multi-version-aware repair prompting, where the model is given information about the target Qiskit version and relevant API migration guides. Automated oracle validation as part of the APR pipeline, and extension of this evaluation protocol to other quantum frameworks, are also natural next steps.

\bibliographystyle{IEEEtran}
\bibliography{references}

\end{document}